\documentclass[aps,prd,preprint,superscriptaddress,tightenlines,nofootinbib]{revtex4}

\usepackage{graphicx}
\usepackage{dcolumn}
\usepackage{bm}

\begin{document}

\preprint{CLNS 04/1871}       
\preprint{CLEO 04-5}         


\title{First Observation and Dalitz Analysis of the $D^0 \to K^0_S \eta \pi^0$ Decay}




\author{P.~Rubin}
\affiliation{George Mason University, Fairfax, Virginia 22030}
\author{C.~ Cawlfield}
\author{B.~I.~Eisenstein}
\author{G.~D.~Gollin}
\author{I.~Karliner}
\author{N.~Lowrey}
\author{P.~Naik}
\author{C.~Sedlack}
\author{M.~Selen}
\author{J.~J.~Thaler}
\author{J.~Williams}
\affiliation{University of Illinois, Urbana-Champaign, Illinois 61801}
% ...
\collaboration{CLEO Collaboration} 
\noaffiliation


\date{September 7, 2004}

\newcommand{\RESRAT}{
     \frac{BR(D^0 \to K^0_S \eta \pi^0)} {BR(D^0 \to K^0_S \pi^0)}
           = 0.46 \pm 0.07 \pm 0.06
                    }

\newcommand{\RESBRVAL}{
            (1.05 \pm 0.16 
                 \pm 0.14 
                 \pm 0.10)\%
                      }
\newcommand{\RESBR}{
     BR(D^0 \to \overline{K}^0 \eta \pi^0) 
         =  \RESBRVAL
                   }

\begin{abstract} 
Using 9.0~fb$^{-1}$ of integrated luminosity in $e^+e^-$ collisions 
near the $\Upsilon$(4S) mass collected with the CLEO~II.V detector 
we report the first observation of the decay $D^0 \to K^0_S \eta \pi^0$.
We measure the ratio of branching fractions, $\RESRAT$.  
We perform a Dalitz analysis of 155 selected $D^0 \to K^0_S \eta \pi^0$ 
candidates and find leading contributions from
$a_0(980) K^0_S$ and $K^*(892) \eta$ intermediate states. 
\end{abstract}

\maketitle

$ $

$ $

A large fraction of the known $D$ meson decay rate is in
three-body hadronic decays to the pseudoscalar particles $K$ and $\pi$.
These decays dominantly proceed through quasi-two-body intermediate
states with a rich set of resonances.
The dynamics of three body decays can be studied using the Dalitz technique~\cite{dalitz}. 
Interest in the decay $D^0 \to K^0_S\eta\pi^0$ stems from
comparing the results of the Dalitz plot analyses 
of the decay $D^0 \to K^0_S\pi^+\pi^-$ studied by 
ARGUS\cite{ARGUS-1993} and CLEO\cite{Asner}
with the decay $D^0 \to K^0_SK^+K^-$ 
studied by ARGUS\cite{ARGUS-1987} and BaBar\cite{BABAR}.
The contribution of the $f_0(980)$ observed in the former case
is not enough to explain the $\sim$60\% fraction observed in the latter decay. 
Additional scalar contribution from $a_0(980)K^0_S$ can be expected in 
$D^0 \to K^0_SK^+K^-$, but is difficult to separate from $f_0(980)K^0_S$
in the Dalitz plot.
The $a_0(980)K^0_S$ intermediate state can also be observed in the favored
$a_0(980)\to \eta \pi^0$ decay mode which would give rise to the 
$D^0 \to K^0_S\eta\pi^0$ final state.  The decay $D^0 \to K^0_S\eta\pi^0$
or any other $D^0$ modes with $a_0(980)$ in the intermediate state
have not yet been observed.  
There is little information on $D^0$ decay modes with $\eta$ in the final state;
only an upper limit $BR(D^0\to \eta X)<13\%$ $@~C.L.=90\%$~\cite{PDG-2002} has been
measured.
Note that $K^0_S\eta\pi^0$ is a CP eigenstate.
A large sample with a good signal to noise ratio in this mode
can be used for studies of CP violation in $D^0$
and $\overline{D^0}$ decays.

The data sample used in this analysis was produced 
by the Cornell Electron Storage Ring (CESR) and
collected with the general purpose CLEO~II.V \cite{Kubota:1992ww}
detector.
Our analysis is based on 9.0~fb$^{-1}$ of integrated luminosity of 
$e^+e^-$ collisions at $\sqrt{s} \simeq 10$~GeV above and below $B\bar{B}$
production threshold.
Charmed particles can be produced both in the process $e^+e^- \to c\bar{c}$
and in $B$ meson decays. 
To suppress events with
low momentum $D^0$'s from $B$ decays, which have higher multiplicity
and higher combinatorial backgrounds, we use the decay 
$D^{*+} \to D^0 \pi^+$ 
(charge conjugation is implied throughout this letter)
as a tag and require that the $D^{*+}$ momentum exceeds 2.8~GeV/c.
The decay $D^0 \to K^0_S \eta \pi^0$ 
is observed in the most probable mode of the final state, 
$K^0_S \to \pi^+ \pi^-$, $\eta \to \gamma \gamma$, $\pi^0 \to \gamma \gamma$.


Charged tracks are required to be well measured in the tracking detectors.
Candidate $K_S^0$-s are reconstructed from pairs of oppositely charged tracks
assumed to be pions. The candidate $K_S^0$ trajectory is required to be 
consistent with production in the interaction region, while its vertex should
be significantly ($>10\sigma$) isolated from this region.
We select $K_S^0$ candidates if the reconstructed mass, 
$m_{\pi^+\pi^-}$, is within 10~MeV/c$^2$ of the nominal 
$K_S^0$ mass \cite{PDG-2002}.
On average, $K^0_S$-s in this selection have a
mass resolution of $\sigma_{K^0_S}=3.7 \pm 0.2~\rm MeV/c^2$.

We form $\pi^0$ and $\eta$ candidates from pairs of neutral showers in the 
CLEO CsI calorimeter.  They are required to be consistent with
electromagnetic showers, have an energy deposition above 30 MeV
and be in the central, barrel region of the detector.  
For $\pi^0$ candidates we require the invariant mass, $m_{\gamma\gamma}$, 
of the photon pair
to be within 18~MeV/c$^2$ of the nominal
$\pi^0$ mass \cite{PDG-2002}.  
The average detector resolution of $\pi^0 \to \gamma\gamma$ invariant mass
is $\sigma_{\pi^0}=6.1\pm1.2~\rm MeV/c^2$.
Similarly, $\eta$ candidates are required to have a two photon 
invariant mass within 40~MeV/c$^2$ of the nominal $\eta$ mass \cite{PDG-2002}
at the average detector resolution of $\sigma_\eta=12.6\pm1.0~\rm MeV/c^2$.

We kinematically fit $K^0_S$, $\pi^0$ and $\eta$ candidates 
and constrain their masses to nominal values.  
This procedure improves the $D^0$ mass resolution by a 
factor of two for $D^0 \to K^0_S\eta\pi^0$ decays.
We reconstruct $D^0\to K^0_S\eta\pi^0$ candidates by combining
the $K_S^0$, $\pi^0$ and $\eta$ candidates in the event.
To eliminate the significant
combinatoric background, we select the combination with the smallest  
\\
$   \chi^2_m = 
              \Big( \frac{m_{\gamma \gamma} - m_\eta}{\sigma_\eta}       \Big)^2 
            + \Big( \frac{m_{\gamma \gamma} - m_{\pi^0}}{\sigma_{\pi^0}} \Big)^2 
            + \Big( \frac{m_{\pi^+ \pi^-} - m_{K^0_S}}{\sigma_{K^0_S}}   \Big)^2,
$
where all the invariant masses are taken before the mass constraint of the kinematic fit.

\begin{figure}
  \includegraphics*[width=3.4in]{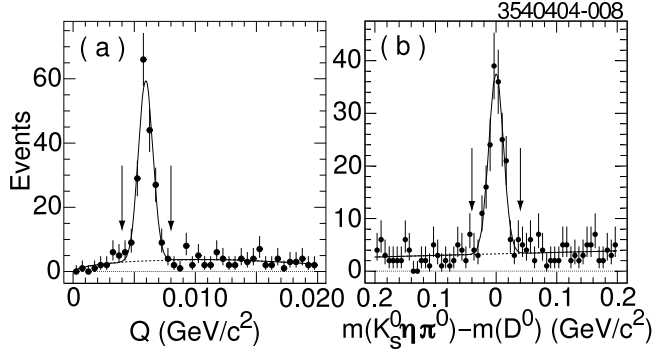}
  \caption{Distribution of the energy release, $Q$, in the decay $D^{*+} \to D^0 \pi^+$ (a), 
   and the mass difference of $D^0 \to K^0_S\eta\pi^0$ candidates (b).}
  \label{fig:ksetapi_dmd0_gp_data}
\end{figure}

The $D^0$ candidate is combined with $\pi^+$ tracks to 
form the tagging decay $D^{*+} \to D^0 \pi^+$.  
A significant $D^0$ signal is observed both in the energy release,
$Q=m(K^0_S\eta\pi^0\pi^+) - m(K^0_S\eta\pi^0) - m_{\pi^+}$, 
and in the $D^0$ mass difference $\Delta m = m(K^0_S\eta\pi^0) - m_{D^0}$ shown in
Fig.~\ref{fig:ksetapi_dmd0_gp_data}.
The $Q$ distribution, shown in Fig.~\ref{fig:ksetapi_dmd0_gp_data}a,
represents raw $Q$ vs $\Delta m$ events
in $\sim 3\sigma$ signal $\Delta m$ band indicated by arrows in 
Fig.~\ref{fig:ksetapi_dmd0_gp_data}b and vice versa.

We estimate a signal yield of 155$\pm$22 events from
a fit with a single Gaussian for the signal plus a linear background 
to the mass spectrum of Fig.~\ref{fig:ksetapi_dmd0_gp_data}(b).
The GEANT-based Monte Carlo simulation \cite{GEANT} of the CLEO II.V
detector response is used to estimate the efficiency 
$\varepsilon (D^0 \to K^0_S \eta \pi^0) = 
(1.15 \pm 0.05 \pm 0.12 \pm 0.01)\%,$
where the uncertainties are  statistical, systematic and 
from the uncertainties on the  
$K^0_S \to \pi^+ \pi^-$,
$\eta \to \gamma \gamma$,
$\pi^0 \to \gamma \gamma$ branching fractions, respectively.
The systematic uncertainty includes 
the track reconstruction efficiency (2\%/track),
$\pi^0$ and $\eta$ selection (5\% each), and the background subtraction in
$D^0$ mass spectrum (7.2\%).  
The first two uncertainties absorb a variation of
efficiency between the phase space and the resonant event 
production mechanism.
The background subtraction error is estimated from
variation in the signal yield when we change the fit function including 
a single versus double Gaussian for the signal and background
described with a linear function, taken from the
$D^0$ mass spectrum sidebands, or taken from the Q distribution sidebands.

To measure the branching fraction we
normalize to the total number of $D^0$'s produced in the decay $D^{*+} \to D^0 \pi^+$.
We use the $D^0 \to K^0_S \pi^0$ decay with known rate,
$BR(D^0 \to K_S^0 \pi^0) = \frac{1}{2} BR(D^0 \to \overline{K}^0 \pi^0) = (1.14 \pm 0.11)\%$
\cite{PDG-2002}.
We use the same selection as $D^0 \to K^0_S \eta \pi^0$, 
but without the $\eta$, and find a very clean $D^0 \to K^0_S \pi^0$
signal with yield of 1105$\pm$54 events and an efficiency
$\varepsilon (D^0 \to K^0_S \pi^0) = (3.76 \pm 0.18 \pm 0.26 \pm 0.02) \%.$
We find the ratio of branching fractions to be
$
\frac{BR(D^0 \to \overline{K}^0 \eta \pi^0)}
     {BR(D^0 \to \overline{K}^0 \pi^0)} =
\frac{BR(D^0 \to K^0_S \eta \pi^0)}
     {BR(D^0 \to K^0_S \pi^0)}
= 0.46  \pm 0.07
        \pm 0.06
        \pm 0.003
= 0.46  \pm 0.09,
$
where the errors are statistical, systematic, and 
$\eta \to \gamma \gamma$ branching fraction uncertainties, respectively.
Using the known $D^0 \to K_S^0 \pi^0$ branching fraction, we
find
$
BR(D^0 \to \overline{K}^0 \eta \pi^0)=
\label{eqn:br_res}
    \RESBRVAL,
$
where the last error is associated with the uncertainty on
the $D^0 \to K^0_S \pi^0$ branching fraction.  Many
systematic uncertainties cancel in the ratio measurement.

\begin{figure}
  \includegraphics*[width=3.4in]{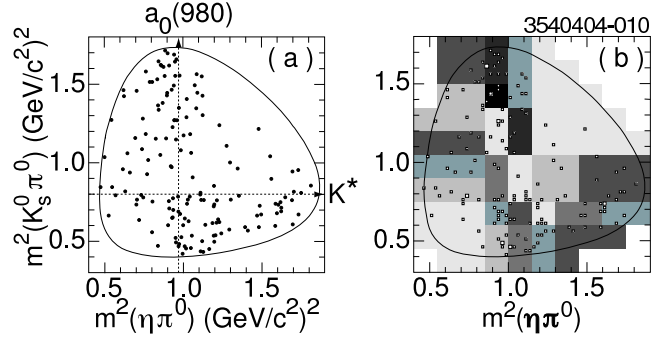}
  \caption{Dalitz plot of $D^0 \to K^0_S\eta\pi^0$ (a), 
           and the map for the adaptive binning (b).}
  \label{fig:ksetapi_dalitz_dots_data}
\end{figure}

The selected sample, although small, is clean enough
to search for possible intermediate states using the Dalitz technique \cite{Bergfeld}.
We tighten the mass difference selection criteria to two standard deviations
($|\Delta m|<25~\rm MeV/c^2$ and $|\Delta Q|<1.2~\rm MeV/c^2$)
in order to increase signal to background ratio.
We select for Dalitz analysis 155 events 
(accidentally the same number of events that 
we find for measurement of the branching ratio)
shown in Fig.~\ref{fig:ksetapi_dalitz_dots_data}(a) as
$m^2(\eta \pi^0)$ versus $m^2(K^0_S \pi^0)$.
The same selection criteria were applied to 
measure the efficiency across the Dalitz plot with a simulation 
of $D^0\to K^0_S\eta\pi^0$ decaying uniformly in its allowed phase space.
The shape of the small background is taken from the data sample of 171 events
in a $Q$ sideband, $10<Q<25~\rm MeV/c^2$, and an
extended range of invariant mass, $|\Delta m|<100~\rm MeV/c^2$. 
Both the efficiency and the background are nearly uniform across 
the Dalitz plot and we parameterize them separately with a two-dimensional 
polynomial of third degree obtained from the dedicated fit.

\begin{figure}
  \includegraphics*[width=3.4in]{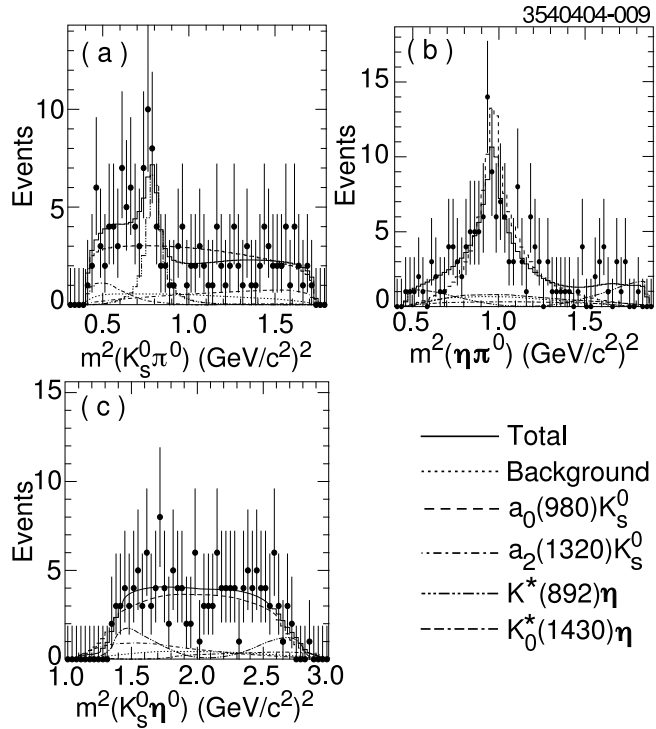}
  \caption{The three projections of the Dalitz plot.
           The fit shown has contributions from $a_0(980) K^0_S$, $K^*(892) \eta$, $K^*_0(1430)\eta$,
           and $a_2(1320)K^0_S$.}
  \label{fig:ksetapi_dp_fract_proj_3}
\end{figure}

The Dalitz plot, Fig.~\ref{fig:ksetapi_dalitz_dots_data}(a),
shows a significant contribution from $a_0(980)K^0_S$ 
interfering with other resonances, as evidenced by
the deficit of
event density in the center of the  plot,  and by the shift
to the left (right) of the $a_0(980)$ band on
the top (bottom) of the plot.
There is an indication of a $K^*(892)\eta$ 
contribution as there is an enhancement in
the expected region in the $m^2(K^0_S \pi^0)$ projection, shown in
Fig.~\ref{fig:ksetapi_dp_fract_proj_3}(a).
The visible mass peak is shifted lower than would
be expected given the $K^*(892)$ mass
indicating interference of $K^*(892) \eta$ with other
intermediate states.

To extract information from the Dalitz plot we apply 
the technique developed in our previous analyses
\cite{Bergfeld}, 
\cite{Asner} 
which uses an unbinned 
maximum likelihood fit and an ``isobar model'' to
measure matrix element amplitudes. 
An isobar model approximates the matrix elements as 
\\
$   {\mathcal M}  = a_{_{NR}} e^{i\varphi_{_{NR}}} 
                 + \sum_{_{R}} a_R e^{i\varphi_{_R}} {\mathcal A}_J(\{K^0_S, \eta, \pi^0\}|R),$
\\
a coherent sum of non-resonant $(NR)$ and resonance $(R)$ terms, 
each multiplied by its own complex factor.
The complex factor is parametrized by a real amplitude $a_{_R}$ and a phase $\varphi_{_R}$, which are extracted from the fit.
The amplitude, ${\mathcal A}_J(ABC|R)$, 
is defined for the decay chain
$D^0 \to RC \to ABC$ with an intermediate resonance $R$
represented by the Breit-Wigner function with spin $J$ dependent factor. 
The overall amplitude normalization and complex phase are arbitrary, 
and are chosen such that $a_{a_0}=1$, $\varphi_{a_0}=0$.

The mass $m$ dependent width of the $a_0(980)$
is parameterized using the method of Flatte~\cite{Flatte},
while the partial width is proportional to the phase space factor $\rho=2p/m$ 
instead of the decay momentum $p$, \\
$ m\Gamma_{a_0(980)} (m) =
\frac{g^2_{a_0\eta \pi^0}}{16\pi} \rho_{\eta \pi^0} +
\frac{g^2_{a_0K^+K^-}}{16\pi}\Big( \rho_{K^+K^-} + \rho_{K^0\bar{K^0}} \Big).
$ 
We assume an isospin symmetry for the coupling constants, 
$g^2_{a_0\eta \pi^+} = g^2_{a_0\eta \pi^0}$ and
$g^2_{a_0K^+K^-} = g^2_{a_0K^0\bar{K^0}} = g^2_{a_0K^0K^+} / 2$.
In our standard fit we use $a_0(980)$ parameters from~\cite{BUGG-1994},
$m(a_0(980))=999 \pm 5~\rm MeV/c^2$,
$g^2_{a_0\eta\pi^0}=11.1 \pm 1.0~\rm GeV^2$,
$g^2_{a_0K^+K^-}/g^2_{a_0\eta\pi^0}=0.58\pm0.09$.

The event density of the Dalitz plot is fit to the efficiency corrected 
matrix element squared and the background polynomial which
is added incoherently \cite{Bergfeld} to the signal.  The relative signal
fraction, $0.867\pm0.027$, is estimated from the $\Delta m$ spectrum
of the data sample.  In all Dalitz fits the signal fraction is
a parameter of the fit constrained to this estimate.

With our
sample we find the most reliable goodness of fit estimator to be
a $\chi^2$-like parameter for Poisson statistics:
$ \chi^2 = \sum_{i=1}^{N} \frac{(n_i-\nu_i)^2}{\nu_i},$~\cite{PDG-2002}
where $n_i$, $\nu_i$ are the number of events and its mean expectation 
in the i-th bin, and $N$ is a total number of bins.
We split the Dalitz plot into $10\times10$ equal bins.
In order to provide sufficient statistics for a mean expectation, it was
necessary to join some bins using so-called ``adaptive binning'' and
requiring $\nu_i$ (or $n_i$)$>$5 in each bin.
The 24 bins found with adaptive binning are shown in
Fig.~\ref{fig:ksetapi_dalitz_dots_data}(b).
We have tested this goodness of fit parameter in
simplified Monte-Carlo based simulations of our data and find
that it gives a uniform probability for statistically distributed data with
$P(\chi^2/N_{d.o.f.})$ in the range [0,1].
The simulation of different models shows
that we are only sensitive to contributions to the Dalitz plot
that are greater than 20\% of the total rate, 
and thus our goal is to find a consistent 
description of the observed event density using a minimal set of 
dominant modes.

From previous observations~\cite{PDG-2002} we expect
$\eta \pi^0$ to have contributions from intermediate states
including 
$a_0(980)$,
$a_2(1320)$, and
$a_0(1450)$.
Similarly, $K^0_S \pi^0$ should have contributions from
$K^*(892)$,
$K^*_1(1410)$,
$K^*_0(1430)$,
$K^*_2(1430)$, and
$K^*_1(1680)$.
A possible low-mass $K\pi$-scalar state or dynamical structure,
$\kappa$, which is not included in \cite{PDG-2002} but is widely discussed
in recent publications \cite{E791}, could also contribute.
There is no obvious contribution from $K^0_S \eta$ in this mass range.
We start with a minimal set of resonances and recognize an additional
resonance as contributing if the fit probability improves, the amplitude
is at least three standard deviations from zero, and the error on the
phase is less than $30^\circ$.

We find that a model including only 
$a_0(980) K^0_S$ and $K^*(892)\eta$ contributions gives a low probability of 
0.8\% and is an unlikely explanation of our data.
Models with a single resonance are even worse with probabilities
of less than $10^{-6}$.
Good consistency with our data can be achieved
with models including two main intermediate states,
$a_0(980) K^0_S$, $K^*(892) \eta$, and additional mode(s).
We find four additional modes giving a fit probability $>$1\%:
(i) a non-resonant fraction;
(ii)  $K_0^*(1430)\eta$;
(iii) $K_0^*(1430)\eta$ and $a_2(1320)K^0_S$ 
      (fit projections are shown in Fig.~\ref{fig:ksetapi_dp_fract_proj_3}); and
(iv) a $\kappa$ with parameters taken from \cite{E791}.
We do not find any significant contribution or fit quality improvement by adding 
other resonances. For these four models
Table~\ref{tab:results_mean} summarizes the amplitude and phase
we extract from the fit for the $K^*(892) \eta$ mode, fixing
the amplitude and phase for the $a_0(980) K^0_S$ mode to be one and zero
respectively.  
Our sample is too small to allow us to choose one model among these four.  
In the last row of Table~\ref{tab:results_mean}
we present averaged results and their variation due to our
inability to choose a single decay model that describes
our data adequately.
\begin{table*}[!htb]
\caption{\label{tab:results_mean} Results for four models of the additional
         contribution beyond $a_0(980) K^0_S$ and 
         $K^*(892) \eta$ to $D^0 \to K^0_S \eta \pi^0$.
         The amplitude and phase for $a_0(980) K^0_S$ are fixed to 1 and $0^\circ$ respectively.
         The uncertainties are statistical from the fit. 
         ``FF(Add.)'' means the sum of the fit fractions for all modes in addition
         to $a_0(980) K^0_S$ and $K^*(892) \eta$ in the model.
         The last row shows averaged values with statistical uncertainties and 
         half the range among the four decay models.
         }
\begin{center}
\begin{tabular}{|c|c|c|c|c|c|c|}
\hline
Additional Mode(s)        & $a_{K^*(892)\eta}$ 
                          & $\varphi_{K^*(892)\eta}({}^\circ)$ 
                          & FF($a_0(980)K^0_S)$  
                          & FF($K^*(892) \eta$)   
                          & FF(Add.) & Prob.,\%\\ 
\hline
NR                        & .234$\pm$.035  
                          & 260$\pm$10 
                          & 1.350$\pm$0.097  
                          & .301$\pm$.071   
                          & .288$\pm$.113
                          & 6.4\\
$K^*_0(1430)\eta$         & .237$\pm$.032  
                          & 258$\pm$10 
                          & 1.322$\pm$0.070  
                          & .301$\pm$.070   
                          & .360$\pm$.115 
                          & 19.4 \\
$K^*_0(1430)\eta+a_2(1320)K^0_S$        
                          & .253$\pm$.031  
                          & 251$\pm$15 
                          & 1.042$\pm$0.146    
                          & .273$\pm$.050   
                          & .316$\pm$.097 
                          & 64.7 \\
$\kappa\eta$              & .269$\pm$.032  
                          & 262$\pm$11 
                          & 1.050$\pm$0.060  
                          & .310$\pm$.060   
                          & .186$\pm$.056 
                          & 49.1 \\ 
\hline
Average and \{Variation\}   
                          & .249$\pm$.032     \{.018\} 
                          & 259$\pm$12        \{6\}
                          & 1.187$\pm$0.093   \{0.154\}
                          & .293$\pm$.062     \{.019\}
                          & .246$\pm$.092     \{.087\}
                          & $-$ \\
\hline
\end{tabular}
\end{center}
\end{table*}

When the amplitudes and phases are extracted from the fit we derive
the fit fraction (FF) for each contribution.
The fit fraction is defined for each resonance
as its matrix element amplitude squared (rate)   
integrated over the allowed phase space 
divided by the total matrix element amplitude squared 
integrated over the same phase space.  
In general the sum of the fit fractions does not
have to equal one due to interference among the contributions.
A statistical uncertainty on the fit fraction is computed 
from the fit covariance matrix
using Monte Carlo methods as described in~\cite{Bergfeld}.  
Table~\ref{tab:results_mean}
gives the fit fractions for the $a_0(980) K^0_S$ and $K^*(892)\eta$
modes, and the fit fraction for the additional mode(s), their 
averaged values, and estimated variations due to the choice of decay model. 

We consider possible sources of systematic uncertainties due to
the background, 
the efficiency, the finite detector resolution, 
the parameterization of the matrix element amplitude,
and choice of decay model. 
Central values are taken as the statistical weighted mean of
the results summarized in Table~\ref{tab:results_mean}.

For the background and the efficiency we perform the fit
with the two-dimensional polynomial coefficients allowed to float
constrained by their covariance matrices.
We also hold the efficiency constant across the Dalitz plot.
Deviations from the standard fit are treated as systematic uncertainties.
The effects are small.

As a consistency check, we allow the parameters of one of the 
clearly observed resonances to float and extract values from the fit.
For both the $K^*(892)$ and $a_0(980)$ we obtain masses,
and width parameters consistent with previously measured values.

Our mass resolution, small compared to the widths of the resonances we          
are considering, is a negligible effect as we observe no change when            
we do a fit that smears each resonance by a two dimensional Gaussian            
with widths given by propagating uncertainties on track fits and shower        
reconstructions.                                                                
 
We also consider variations in the description of the
decay amplitudes.  We vary the radial parameters for the intermediate resonances
between zero and twice their standard value of $\sim3~\rm GeV^{-1}$
\cite{Bergfeld}.
We allow the masses and widths
for the intermediate resonances to vary within one standard deviation of
their measured values~\cite{PDG-2002}. 
The largest variation from the standard fit of each fit parameter is
taken as an uncertainty.
These uncertainties are combined quadratically to give a
systematic uncertainty.

The largest systematic uncertainty results from choice of
decay model.  Using the four models giving good fits
we take half the range of central values, shown
in Table~\ref{tab:results_mean},
as this uncertainty and report it separately.


Our analysis apparently contradicts a result done with an earlier
version of our detector 
$BR(D^0 \to \overline{K}^*(892)\eta)= (1.8 \pm 0.4)$\%
\cite*{CLEO_KIN91,CLEO_PRO93,PDG-2002}.
That analysis, which focused on a search for this mode, made helicity
angle and $\eta$ momentum selections that are 
not compatible with a $K^0_S \eta \pi^0$ Dalitz analysis.
Thus the effects of interference were not considered. 
Comparing the fit result to the $K\pi$ mass spectrum in \cite{CLEO_PRO93}
with results obtained in this analysis we find that the
$D^0 \to \overline{K}^*(892)\eta$ rate is larger by
roughly a factor of two. 

In conclusion, we have observed for the first time the decay 
$D^0 \to K^0_S \eta \pi^0$.
We have measured the ratio of the branching fractions,
\begin{equation}
\label{eqn:br_ratio_sum}
      \RESRAT ,
\end{equation}
where the uncertainties are statistical and systematic
respectively.
Using the known $D^0 \to K^0_S \pi^0 (\overline{K^0} \pi^0)$
decay rate we measure the branching fraction
\begin{equation}
\label{eqn:br_sum}
    \RESBR,
\end{equation}
where the final uncertainty is associated  
with the $D^0 \to \overline{K}^0 \pi^0$ branching fraction.

We have analyzed the resonant substructure of the decay $D^0 \to K^0_S \eta \pi^0$
using the Dalitz technique.
We find dominant contributions from $a_0(980)K^0_S$ and $K^*(892) \eta$
intermediate states. 
Using an isobar model including $K^*(892)\eta$, $a_0(980) K^0_S$, 
and averaging over four consistent models for
additional components we find the amplitude, phase and fit fractions \\
\centerline{
$a_{K^*(892)\eta} = 0.249 \pm 0.032 \pm 0.013 \pm 0.018, $ }
\centerline{$\varphi_{K^*(892)\eta} = (259 \pm 12 \pm 9 \pm 6)^\circ, $}
\centerline{$FF(K^*(892)\eta) = 0.293 \pm 0.062 \pm 0.029 \pm 0.019, $}
\begin{equation}
FF(a_0(980) K^0_S) = 1.19 \pm 0.09 \pm 0.20 \pm 0.16,
\end{equation}
where $a_{a_0K^0_S}$ and $\varphi_{a_0K^0_S}$ are
fixed to one and zero respectively.  
The uncertainties are
statistical, systematic, and decay model choice respectively.
We also find that contributions from $a_0(980) K^0_S$ and $K^*(892) \eta$
are not sufficient to describe our data.
We estimate the fit fraction of any additional component as
\begin{equation}
FF(\rm Add.) = 0.246\pm0.092\pm 0.025 \pm 0.087,
\end{equation}
with the uncertainties meaning as above.

We gratefully acknowledge the effort of the CESR staff
in providing us with
excellent luminosity and running conditions.
This work was supported by
the National Science Foundation and
the U.S. Department of Energy.

\end{document}